\documentclass[a4paper,11pt]{article}

\usepackage{graphicx}

\usepackage{amsmath}
\usepackage{amsfonts}
\usepackage{amssymb}
\usepackage{latexsym}
\usepackage{amsthm}  

\usepackage[textwidth=15cm,textheight=24cm]{geometry}

\begin{document}

\title{An efficient approximation for point-set diameter \\ in higher dimensions}
\author{Mahdi Imanparast\footnote{Department of Mathematics and Computer Science, Amirkabir University of Technology, Tehran, Iran}, \hspace*{0.2cm}Seyed Naser Hashemi\footnotemark[1], \hspace*{0.2cm}Ali Mohades\footnotemark[1] \footnote{Laboratory of Algorithms and Computational Geometry, Amirkabir University of Technology, Tehran, Iran}}

\date{}
\maketitle

\begin{abstract}
In this paper, we study the problem of computing the diameter of a  set of $n$ points in $d$-dimensional Euclidean space for a fixed dimension $d$, and propose a new $(1+\varepsilon)$-approximation algorithm with $O(n+ 1/\varepsilon^{d-1})$ time and $O(n)$ space, where $0 < \varepsilon\leqslant 1$. We also show that the proposed algorithm can be modified to a $(1+O(\varepsilon))$-approximation algorithm with $O(n+ 1/\varepsilon^{\frac{2d}{3}-\frac{1}{3}})$ running time. These results provide some improvements in comparison with existing algorithms in terms of simplicity and data structure. \\

% %\hspace*{-0.5cm}\textbf{Keywords} diameter, point set, % %approximation algorithm, higher dimensions.   
\end{abstract}

\section{Introduction}
Given a finite set $\mathcal{S}$ of $n$ points, the diameter of $\mathcal{S}$, denoted by $D(\mathcal{S})$ is the maximum distance between two points of $\mathcal{S}$. Namely, we want to find a diametrical pair $p$ and $q$ such that $D(\mathcal{S})=\max_{\forall{p,q\in \mathcal{S}}} (||p-q||)$. Computing the diameter of a set of points has a large history, and it may be required in various fields such as database, data mining, and vision. A trivial brute-force algorithm for this problem is as follows: compute the distance between each pair of points and then choose the maximum distance. Since computing the distance takes constant time, this algorithm takes $O(n^{2})$ time, but this is too slow for large-scale datasets that occur in the fields. Hence, we need a faster algorithm which may be exact or is an approximation. 

By reducting from the set disjointness problem, it can be shown that computing the diameter of $n$ points in $\mathbb{R}^{d}$ requires $\Omega(n\log n)$ operations in the algebraic computation-tree model~\cite{Prep85}. It is shown by Yao that it is possible to compute the diameter in sub-quadratic time in each dimension~\cite{Yao82}. There are well-known solutions in two and three dimensions. In the plane, this problem can be computed in optimal time $O(n\log n)$, but in three dimensions, it is more difficult. Clarkson and Shor~\cite{Clar89} present an $O(n \log n)$-time randomized algorithm. Their algorithm needs to compute the intersection of $n$ balls (with the same radius) in $\mathbb{R}^{3}$. It may be slower than the brute-force algorithm for the most practical datasets. Moreover, it is not an efficient method for higher dimensions because the intersection of $n$ balls with the same radius has a large size. Some deterministic algorithms with running time $O(n \log ^{3} n)$~\cite{Ama94,Ram97a} and $O(n \log ^{2} n)$~\cite{Ram97b,Bes01} are found for solving this problem. Finally, Ramos~\cite{Ram00,Ram01} introduced an optimal deterministic $O(n \log n)$-time algorithm in $\mathbb{R}^{3}$. Cheong et al.~\cite{Cheon01} present an $O(n \log^{2} n)$ randomized algorithm that computes the all-pairs farthest neighbors for $n$ points on the convex position in $\mathbb{R}^{3}$. \\
\hspace*{0.5cm}In the absence of fast algorithms, many attempts have been done to approximate the diameter in low and high dimensions. A 2-approximation algorithm with $O(dn)$ time in $d$ dimensions can be found easily by selecting a point $x$ of $\mathcal{S}$ and then finding the farthest point $y$ from it by brute-force manner. The first non-trivial approximation algorithm for the diameter is presented by Egecioglu and Kalantari~\cite{Ege89} that approximates the diameter with factor $\sqrt{3}$ and operations cost $O(dn)$ in $d$ dimensions. They also present an iterative algorithm with $t\leq n$ iterations and the cost $O(dn)$ for each iteration that has approximation factor $\sqrt{5-2\sqrt{3}}$. Agarwal et al.~\cite{AMS92} present a $(1+\varepsilon)$-approximation algorithm in $\mathbb{R}^{d}$ with $O(n/ \varepsilon^{(d-1)/2})$ running time by projection to directions. Barequet and Har Peled~\cite{Bar01} present a $\sqrt{d}$-approximate diameter method with $O(dn)$ time. They also describe a $(1+\varepsilon)$-approximate approach for computing the diameter with $O(n+1/\varepsilon^{2d})$ time in $\mathbb{R}^{d}$. They show that the running time can be improved to $O(n+1/\varepsilon^{2(d-1)})$. Similarly, Har Peled~\cite{HarP01} presents an  approach which for the most inputs 
is able to compute very fast the exact diameter, or an approximation. Although, in the worst case, the algorithm running time is still quadratic, and it is sensitive to the 
hardness of the input. His algorithm is  able to return a pair of points $p$ and $q$ such that $||p-q||\geq (1-\varepsilon)D$, for each value $\varepsilon>0$ in each dimension with $O((n+ 1/\varepsilon^{2d}) \log 1/\varepsilon)$ running time. He shows that with a complicated analysis, this running time can be reduced to $O((n+ 1/\varepsilon^{3(d-1)/2}) \log 1/\varepsilon)$. Simultaneously, Maladain and Boissonnat~\cite{Mala02} present an exact algorithm for the diameter by computing the double normals in each dimension, but their algorithm is not worst-case optimal. They also show that with having double normals, a $\sqrt{3}$-approximation of the diameter in each dimension is provided. Moreover, Finocchiaro and Pellegrini~\cite{Fino02} describe an algorithm that finds in $O(dn \log n+n^{2})$ time with high probability a $(1+\varepsilon)$-approximation for the diameter of a set of $n$ points in $d$-dimensional Euclidean space. Chan~\cite{Cha02} observes that a combination of two approaches in~\cite{AMS92} and~\cite{Bar01} yields a $(1+\varepsilon)$-approximation algorithm with $O(n+1/\varepsilon^{3(d-1)/2})$ time and a $(1+O(\varepsilon))$-approximation algorithm with  $O(n+1/\varepsilon^{d-\frac{1}{2}})$ time. He also introduces a core-set theorem, and shows that using this theorem, a $(1+O(\varepsilon))$-approximation for the diameter in $O(n+1/\varepsilon^{d-\frac{3}{2}})$ time can be found~\cite{Cha06}. Recently, Chan~\cite{Chan17} has proposed an approximation algorithm with $O((n/\sqrt{\varepsilon}+1/\varepsilon^{\frac{d}{2}+1})(\log\frac{1}{\varepsilon})^{O(1)})$ time by applying the Chebyshev polynomials for the diameter in low constant dimensions, and  Arya et al.~\cite{AFM17} show that by applying an efficient decomposition of a convex body using a hierarchy of Macbeath regions, it is possible to compute an approximation for the diameter of a point set in $O(n \log \frac{1}{\varepsilon}+1/\varepsilon^{\frac{(d-1)}{2}+\alpha})$ time, where $\alpha$ is
an arbitrarily small positive constant. Table~\ref{tab:tabe1} provides a summary on some non-constant approximation algorithm for the point-set diameter.

\vspace {-0.3 cm} 
\begin{table}‎
\caption{A summary on the complexity of some utilizable non-constant approximation algorithm for the diameter of a point set in $d$-dimensional Euclidean space. Our results are denoted by +.}‎
\label{tab:tabe1}‎
\vspace {-1.1 cm} 
\begin{center}‎
\begin{tabular}[c]{lccc}
\\
\hline‎
\bf Reference & \bf  \hspace{0.05cm} Approximation Factor & \bf \hspace{0.05cm} Running Time 

\\
\hline‎ 
~\cite{AMS92} & $1+\varepsilon$ &$O(\dfrac{n^{^{}}}{\varepsilon^{(d-1)/2}})$   \\
\hspace*{0.1cm}~\cite{Bar01} & $1+\varepsilon$   &$O(n+1/\varepsilon^{2(d-1)})$  \\‎
~\cite{HarP01}  & $1+\varepsilon$ & $O((n+ 1/\varepsilon^{2d}) \log \frac{1}{\varepsilon})$   \\
\hspace*{0.1cm}~\cite{Cha02}  & $1+\varepsilon$ &  $O(n+1/\varepsilon^{\frac{3(d-1)}{2}})$    \\‎
\hspace*{0.2cm} +  & $1+\varepsilon$  & $O(n+ 1/\varepsilon^{d-1})$  \\
\hspace*{0.1cm}~\cite{Cha02}  & $1+O(\varepsilon)$  & $O(n+1/\varepsilon^{d-\frac{1}{2}})$    \\‎
~\cite{Cha06}  & $1+O(\varepsilon)$   & $O(n+1/\varepsilon^{d-\frac{3}{2}})$  
\\‎
~\cite{Chan17}   & $1+O(\varepsilon)$  & $O(n/\sqrt{\varepsilon} (\log\frac{1}{\varepsilon})^{O(1)}+1/\varepsilon^{\frac{d}{2}+1} (\log\frac{1}{\varepsilon})^{O(1)})$  

\\‎
~\cite{AFM17}  & $1+O(\varepsilon)$ & $O(n \log \frac{1}{\varepsilon}+1/\varepsilon^{\frac{(d-1)}{2}+\alpha})$   
\\‎

\hspace*{0.2cm} +  & $1+O(\varepsilon)$  & $O(n+ 1/\varepsilon^{\frac{2d}{3}-\frac{1}{3}})$ 
\\
\hline‎ 
\end{tabular}‎
\end{center}‎
\end{table}‎

\vspace {0.52 cm}  
\subsection{Our results}
In this paper, we propose a new $(1+\varepsilon)$-approximation algorithm for computing the diameter of a set $\mathcal{S}$ of $n$ points in $\mathbb{R}^{d}$ with $O(n+ 1/\varepsilon^{d-1})$ time and $O(n)$ space, where $0 < \varepsilon\leqslant 1$. Moreover, we show that the proposed algorithm can be modified to a $(1+O(\varepsilon))$-approximation algorithm with $O(n+ 1/\varepsilon^{\frac{2d}{3}-\frac{1}{3}})$ time and $O(n)$ space. As stated above, two new results have been recently presented for the diameter problem in~\cite{Chan17} and~\cite{AFM17}. It should be noted that our algorithms are completely different in terms of computational technique. The polynomial technique provided by Chan~\cite{Chan17} is based on using Chebyshev polynomials and discrete upper envelope subroutine~\cite{Cha06}, and the method presented by Arya et al.~\cite{AFM17} requires the use of complex data structures to approximately answer queries for polytope membership, directional width, and nearest-neighbor. While our algorithms in comparison with these algorithms are simpler in terms of understanding and data structure. The remainder of this paper is organized as follows: in section 2, we describe our proposed algorithm. Subsection 2.1 includes our analysis over the algorithm. Subsection 2.2 presents a modified version of the proposed algorithm. In section 3, we draw our conclusion.
\section{The proposed algorithm}	
\begin{figure}
    \begin{center}
       \includegraphics [height=5.05 cm]{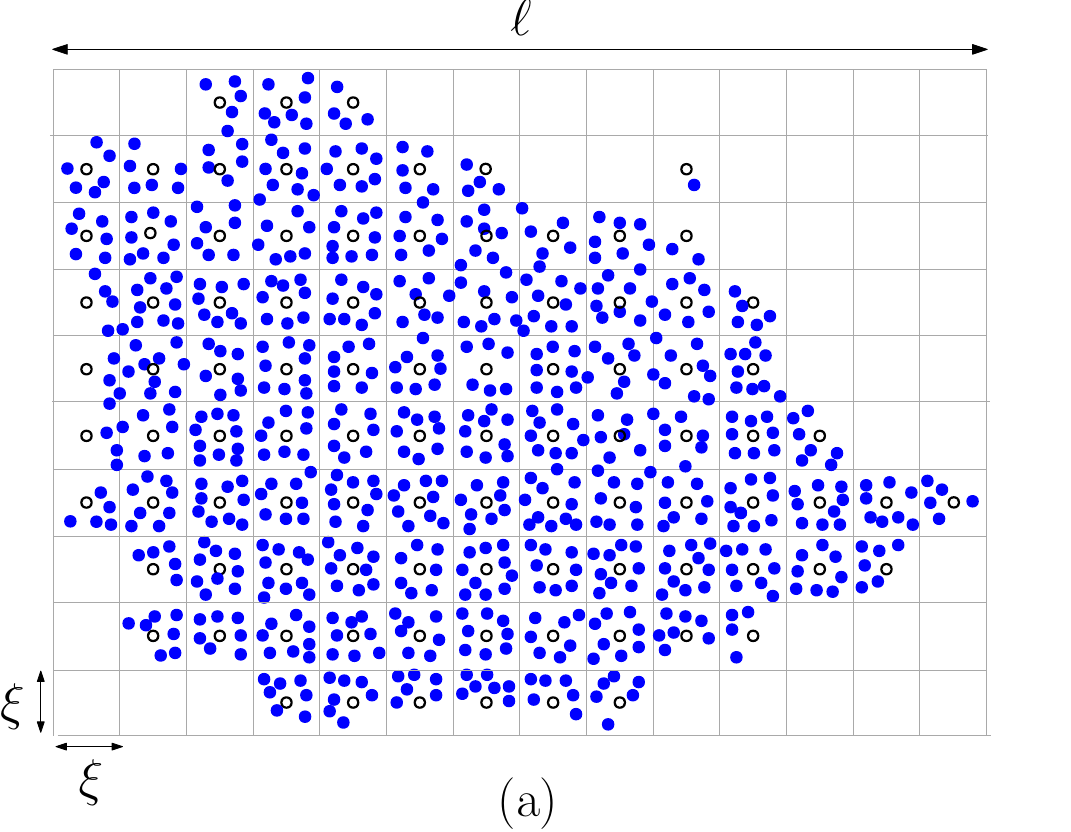}
       \hspace{0.1cm}
       \includegraphics [height=5.05 cm]{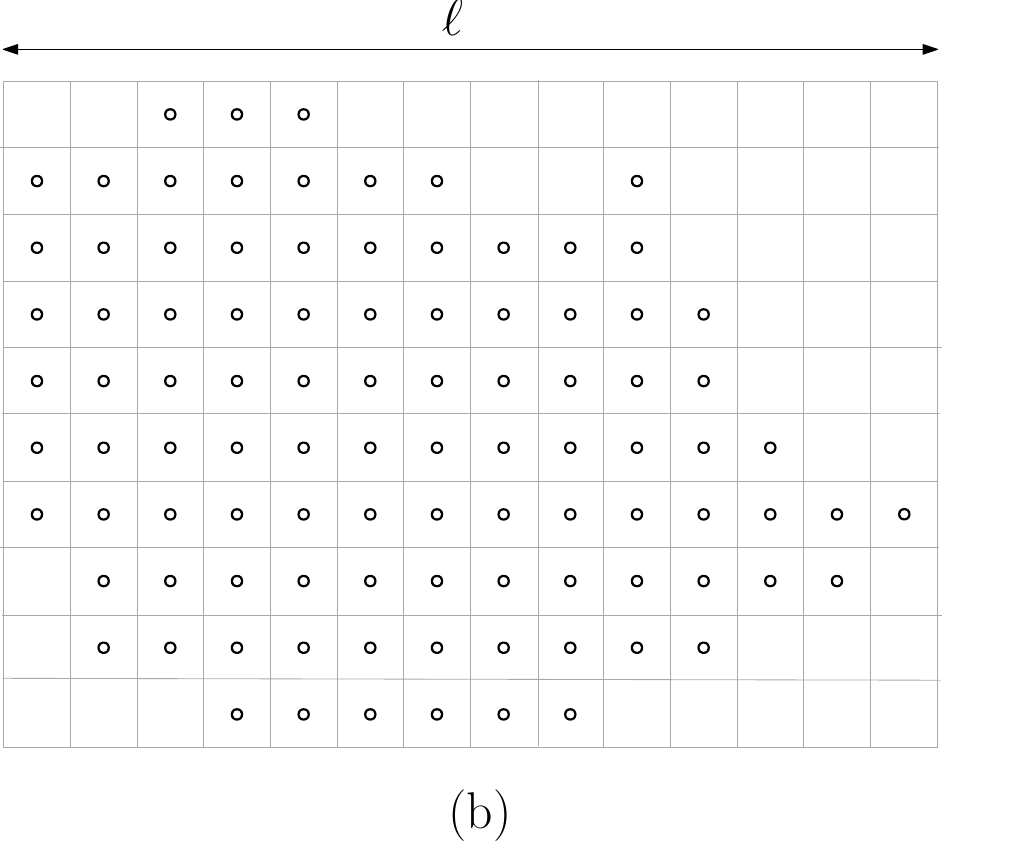}
       \caption{(a) A set of points in $\mathbb{R}^{2}$ and an $\xi$-gird. Initial points are shown by blue points and their corresponding central cell-points are shown by circle points. (b) Rounded point set $\hat{\mathcal{S}}$.}
       \vspace {-0.6 cm}
       \label{fig:Round}            
    \end{center}
\end{figure}

In this section, we describe our new approximation algorithm to compute the diameter of a set $\mathcal{S}$ of $n$ points in $\mathbb{R}^{d}$. In order to follow our algorithm, we first find extreme points in each coordinate and compute the axis-parallel bounding box of $\mathcal{S}$, which is denoted by $B(\mathcal{S})$. We use the largest length side $\ell$ of $B(\mathcal{S})$ to impose grids on the point set. In fact, we first decompose $B(\mathcal{S})$ to a grid of regular hypercubes with side length $\xi$, where $\xi=\varepsilon \ell/2\sqrt{d}$. We call each hypercube a cell. Then, each point of $\mathcal{S}$ is rounded to its corresponding central cell-point. Figure~\ref{fig:Round} shows an example of the rounding process for a point set in $\mathbb{R}^{2}$.\\ 

\vspace {-0.5 cm} 
\vspace {0.5 cm}

\begin{figure}
    \begin{center}
       \includegraphics [height=6.5 cm]{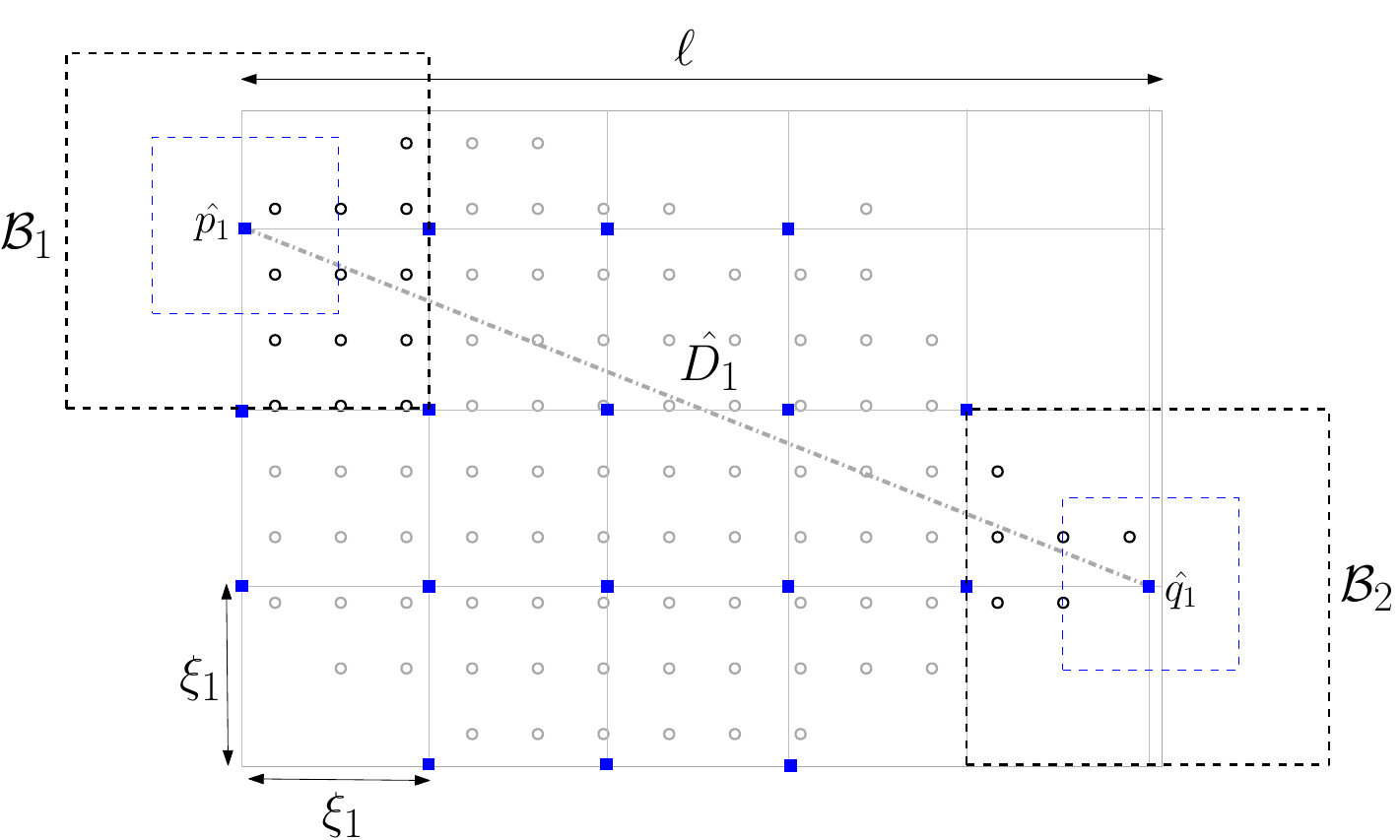}
       \caption{Points of the set $\hat{\mathcal{S}}$ are shown by circle points and their corresponding nearest grid-points in set $\hat{\mathcal{S}_{1}}$ are shown by blue square points. The searching domain for finding the diameter of the point set $\hat{\mathcal{S}}$ is reduced into two point sets $\mathcal{B}_{1}$ and $\mathcal{B}_{2}$.}
       \vspace {-0.7 cm}
      \label{fig:Round2}                   
    \end{center}
\end{figure}

In the following, we impose again a $\xi_{1}$-grid to $B(\mathcal{S})$ for $\xi_{1}=\sqrt{\varepsilon} \ell/2\sqrt{d}$. Then, we round each point of the rounded point set $\hat{\mathcal{S}}$ to its nearest grid-point in this new grid that results in a point set $\hat{\mathcal{S}_{1}}$. Let, $\mathcal{B}_{\delta}(p)$ be a hypercube with side length $\delta$ and central-point $p$. We restrict our search for finding diametrical pairs of the first rounded point set $\hat{\mathcal{S}}$ into two hypercubes $\mathcal{B}_{2\xi_{1}}(\hat{p}_{1})$ and $\mathcal{B}_{2\xi_{1}}(\hat{q}_{1})$ corresponding to two diametrical pair $\hat{p}_{1}$ and $\hat{q}_{1}$ in the point set $\hat{\mathcal{S}_{1}}$. Let us use two point sets $\mathcal{B}_{1}$ and $\mathcal{B}_{2}$ for maintaining points of the rounded point set $\hat{\mathcal{S}}$, which are inside two hypercubes $\mathcal{B}_{2\xi_{1}}(\hat{p}_{1})$ and $\mathcal{B}_{2\xi_{1}}(\hat{q}_{1})$, respectively. See Figure~\ref{fig:Round2}. Then, it is sufficient to find a diameter between points of $\hat{\mathcal{S}}$, which are inside two point sets $\mathcal{B}_{1}$ and $\mathcal{B}_{2}$. We use notation $Diam(\mathcal{B}_{1},\mathcal{B}_{2})$ for the process of computing the diameter of point set $\mathcal{B}_{1}\cup \mathcal{B}_{2}$. Altogether, we can present the following algorithm. 

\vspace*{0.04cm}
\smallskip‎
\hrule‎
\\
\textbf{ Algorithm 1‎:} APPROXIMATE DIAMETER $ (\mathcal{S},\varepsilon)$ 
\hrule‎
\\ 
\footnotesize‎ 
\smallskip‎
\textbf{‎Input‎:} a set $\mathcal{S}$ of $n$ points in $\mathbb{R}^{d}$ and an error parameter $\varepsilon$.

\hspace*{-0.4cm}\textbf{‎Output‎:} approximate diameter $\tilde{D}$.

\hspace{-0.4 cm} 1: \hspace{0.03cm} Compute the axis-parallel bounding box $B(\mathcal{S})$ for a point set $\mathcal{S}$.

\hspace{-0.4 cm} 2: \hspace{0.03cm} $\ell \leftarrow$ Find the length of the largest side in $B(\mathcal{S})$. 

\hspace{-0.4 cm} 3: \hspace{0.03cm} Set $\xi\leftarrow \varepsilon \ell/2\sqrt{d}$ and $\xi_{1}\leftarrow \sqrt{\varepsilon} \ell/2\sqrt{d}$.

\hspace{-0.4 cm} 4: \hspace{0.03cm} $\hat{\mathcal{S}}\leftarrow$ Round each point of $\mathcal{S}$ to its central-cell point in a $\xi$-grid.

\hspace{-0.4 cm} 5: \hspace{0.03cm} $\hat{\mathcal{S}_{1}}\leftarrow$ Round each point of $\hat{\mathcal{S}}$ to its nearest grid-point in a $\xi_{1}$-grid.

\hspace{-0.4 cm} 6: \hspace{0.03cm} $\hat{D}_{1}\leftarrow$ Compute the diameter of the point set $\hat{\mathcal{S}_{1}}$ by brute-force manner, and \\ \hspace*{1.72cm} simultaneously, a list of the diametrical pair $(\hat{p}_{1},\hat{q}_{1})$, such that $\hat{D}_{1}=||\hat{p}_{1}-\hat{q}_{1}||$.

\hspace{-0.4 cm} 7: \hspace{0.03cm} Find points of $\hat{\mathcal{S}}$ which are in two hypercubes $\mathcal{B}_{1}=\mathcal{B}_{2\xi_{1}}(\hat{p}_{1})$ and $\mathcal{B}_{2}=\mathcal{B}_{2\xi_{1}}(\hat{q}_{1})$ \\ \hspace*{0.84cm} for each diametrical pair $(\hat{p}_{1},\hat{q}_{1})$. 
 
\hspace{-0.4 cm} 8: \hspace{0.04cm} $\hat{D}\leftarrow$ Compute $Diam(\mathcal{B}_{1},\mathcal{B}_{2})$, corresponding to each diametrical pair $(\hat{p}_{1},\hat{q}_{1})$ \\ \hspace*{1.57cm} by brute-force manner and return the maximum value between them.

\hspace{-0.4 cm} 9: \hspace{0.03cm} $\tilde{D}\leftarrow\hat{D}+\varepsilon\ell/2$.

\hspace{-0.4 cm} 10: \hspace{0.01cm} Output $\tilde{D}$.
\smallskip‎
\hrule‎
\rmfamily‎ 
\normalsize‎

Note that we only need $O(dn)$ time to round points to their central-cell points. We need $O(d)$ time for computing the central cell-point for each point. Thus, we can round all point of a set of $n$ points to their central-cell points in $O(dn)$ time. Similarly, rounding $n$ points to their nearest grid-point can be done in $O(dn)$ time. 

\vspace {-0.4 cm} 
\vspace {0.2 cm}

\subsection{Analysis}
In this subsection, we analyze the proposed algorithm.

\newtheorem{The1} [enumi] {Theorem}{\bfseries}{\itshape}
\begin{The1}
Algorithm 1 computes an approximate diameter for a set $\mathcal{S}$ of $n$ points in $\mathbb{R}^{d}$ in $O(n+1/\varepsilon^{d-1})$ time and $O(n)$ space, where $0<\varepsilon\leqslant 1$.
\end{The1}
\begin{proof}
Finding the extreme points in all coordinates and finding the largest side of $B(\mathcal{S})$ can be done in $O(dn)$ time. The rounding step takes $O(d)$ time for each point, and for all of them takes $O(dn)$ time, but for computing the diameter over the rounded point set $\hat{\mathcal{S}_{1}}$ we need to know the number of points in the set $\hat{\mathcal{S}_{1}}$. We know that the largest side of the bounding box $B(\mathcal{S})$ has length $\ell$ and the side length of each cell in $\xi_{1}$-grid is $\xi_{1}=\sqrt{\varepsilon} \ell/2\sqrt{d}$. On the other hand, the volume of a hypercube of side length $L$ in $d$-dimensional space is $L^{d}$. Since, corresponding to each point in the point set $\hat{\mathcal{S}_{1}}$, we can take a hypercube of side length $\xi_{1}$. Therefore, the number of grid-points in an imposed $\xi_{1}$-grid to the bounding box $B(\mathcal{S})$ is at most
\begin{equation}
\dfrac{(\ell+\xi_{1})^{d}}{(\xi_{1})^{d}}=\left(\dfrac{\ell}{\sqrt{\varepsilon}\ell/2\sqrt{d}}+1\right)^{d}=\left(\dfrac{2\sqrt{d}} {\sqrt{\varepsilon}}+1\right)^{d}
=O\left(\dfrac{(2\sqrt{d})^{d}}{\varepsilon^{\frac{d}{2}}}\right) .
\end{equation}
So, the number of points in $\hat{\mathcal{S}_{1}}$ is at most $O((2\sqrt{d})^{d} / \varepsilon^{\frac{d}{2}})$. Hence, by the brute-force quadratic algorithm, we need $O((2\sqrt{d})^{d} / \varepsilon^{\frac{d}{2}})^{2})=O((2\sqrt{d})^{2d} / \varepsilon^{d})$ time for computing all distances between grid-points of the set $\hat{\mathcal{S}_{1}}$, and its diametrical pair list. Then, for a diametrical pair $(\hat{p}_{1},\hat{q}_{1})$ in point set $\hat{\mathcal{S}_{1}}$, we compute two sets $\mathcal{B}_{1}$ and $\mathcal{B}_{2}$. They include points of $\hat{\mathcal{S}}$ which are inside two hypercubes $\mathcal{B}_{2\xi_{1}}(\hat{p}_{1})$ and $\mathcal{B}_{2\xi_{1}}(\hat{q}_{1})$, respectively. This work takes $O(dn)$ time. In addition, for computing the diameter of point set $\mathcal{B}_{1}\cup \mathcal{B}_{2}$, we need to know number of points in each of them. On the other hand, the number of points in two sets $\mathcal{B}_{1}$ or $\mathcal{B}_{2}$ is at most
\begin{equation}
\dfrac{Vol(\mathcal{B}_{2\xi_{1}})}{Vol(\mathcal{B}_{\xi})}=\dfrac{(2\sqrt{\varepsilon} \ell/ 2\sqrt{d})^{d}}{(\varepsilon \ell/ 2\sqrt{d})^{d}}=\dfrac{(2\sqrt{\varepsilon})^{d}} {\varepsilon^{d}}
=\dfrac{(2)^{d}}{\varepsilon^{\frac{d}{2}}}.
\end{equation}
Hence, for computing $Diam(\mathcal{B}_{1},\mathcal{B}_{2})$, we need $O(((2)^{d}/\varepsilon^{\frac{d}{2}})^{2})=O((2)^{2d}/\varepsilon^{d})$ time by brute-force manner, but we might have more than one diametrical pair $(\mathcal{B}_{1},\mathcal{B}_{2})$. Since the point set $\hat{\mathcal{S}}_{1}$ is a set of grid-points, so we could have in the worst-case $O(2^{d})$ different diametrical pairs $(\mathcal{B}_{1},\mathcal{B}_{2})$ in point set $\hat{\mathcal{S}}_{1}$. This means that this step takes at most $O(2^{d}\cdot (2)^{2d}/\varepsilon^{d})=O((2\sqrt{2})^{2d}/\varepsilon^{d})$ time. 

\hspace*{-0.5cm}Now, we can present the complexity of our algorithm as follows: 
$$T_{d}(n)=O(dn)+O(dn)+O\left(\dfrac{(2\sqrt{d})^{2d}}{\varepsilon^{d}}\right)+O(2^{d}dn)+O\left(\dfrac{(2\sqrt{2})^{2d}}{\varepsilon^{d}}\right),$$
$$\hspace{0.52cm} \leqslant O(2^{d}dn)+O\left(\dfrac{(2\sqrt{d})^{2d}}{\varepsilon^{d}}\right),$$
\begin{equation}
= O\left(2^{d}dn+ \dfrac{(2\sqrt{d})^{2d}}{\varepsilon^{d}}\right).
\end{equation}
Since $d$ is fixed, we have:
\begin{equation}
T_{d}(n)=O(n+\dfrac{1}{\varepsilon^{d}}).
\end{equation}
We can also reduce the running time of the Algorithm 1 by discarding some internal points which do not have any potential to be the diametrical pairs in rounded point set $\hat{\mathcal{S}_{1}}$, and similarly, in two point sets $\mathcal{B}_{1}$ and $\mathcal{B}_{2}$. This can be done by considering all the points which are same in their $(d-1)$ coordinates and keep only highest and lowest. Then, the number of points in point set $\hat{\mathcal{S}_{1}}$, and two point sets $\mathcal{B}_{1}$ and $\mathcal{B}_{2}$ can be reduced to $O(1/ \varepsilon^{\frac{d}{2}-\frac{1}{2}})$. So, using the brute-force quadratic algorithm, we need $O((1/ \varepsilon^{\frac{d}{2}-\frac{1}{2}})^2)$ time to find the diametrical pairs. Hence, this gives us the total running time $O(n+1/ \varepsilon^{d-1})$.
About the required space, we only need $O(n)$ space for storing required points sets. So, this completes the proof.
\end{proof}

Now, we explain the details of the approximation. 
\newtheorem{The2} [enumi] {Theorem}{\bfseries}{\itshape}
\begin{The2}
Algorithm 1 computes an approximate diameter $\tilde{D}$ such that:
$D \leqslant \tilde{D}\leqslant (1+\varepsilon)D$, where $0 < \varepsilon \leqslant 1$.
\end{The2}
\begin{proof}
In line 7 of the Algorithm 1, we compute two point sets $\mathcal{B}_{1}$ and $\mathcal{B}_{2}$, for each diametrical pair $(\hat{p}_{1},\hat{q}_{1})$ in the point set $\hat{\mathcal{S}_{1}}$. We know that a grid-point $\hat{p}_{1}$ in point set $\hat{\mathcal{S}_{1}}$ is formed from points of the set $\hat{\mathcal{S}}$ which are inside hypercube $B_{\xi_{1}}(\hat{p}_{1})$. We use a hypercube $\mathcal{B}_{1}$ of side length $2\xi_{1}$ to make sure that we do not lost any candidate diametrical pair of the first rounded point set $\hat{\mathcal{S}}$ around a diametrical point $\hat{p}_{1}$, (see Figure~\ref{fig:Round2}). In the next step, we should find the diametrical pair $(\hat{p},\hat{q})\in \hat{\mathcal{S}}$ for points which are inside two point sets $\mathcal{B}_{1}$ and $\mathcal{B}_{2}$. Hence, it is remained to show that the diameter, which is computed by two points $\hat{p}$ and $\hat{q}$, is a $(1+\varepsilon)$-approximation of the true diameter. Let $\hat{p}$ and $\hat{q}$ be two central-cell points of the rounded point set $\hat{\mathcal{S}}$ which are used in line 8 of the Algorithm 1 for computing the approximate diameter $\hat{D}$. Then, we have two cases, either two true points $p$ and $q$ are in farthest distance from each other in their corresponding cells (Figure~\ref{fig:Lem} (a)), or they are in nearest distance from each other (Figure~\ref{fig:Lem} (b)). It is obvious that the other cases are between these two cases.
 
\vspace {-0.1 cm} 
\begin{figure}
    \begin{center}
       \includegraphics [height=4.75 cm]{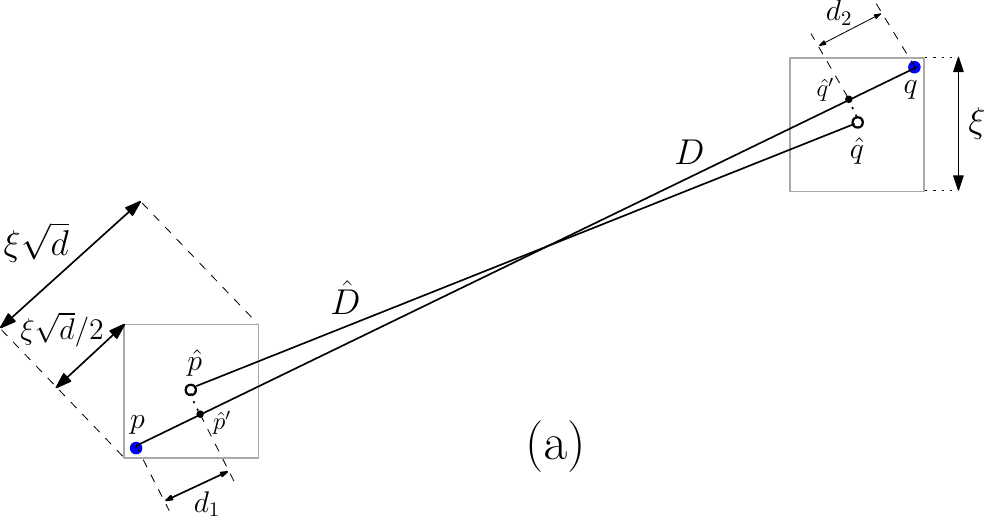}
       \includegraphics [height=3.85 cm]{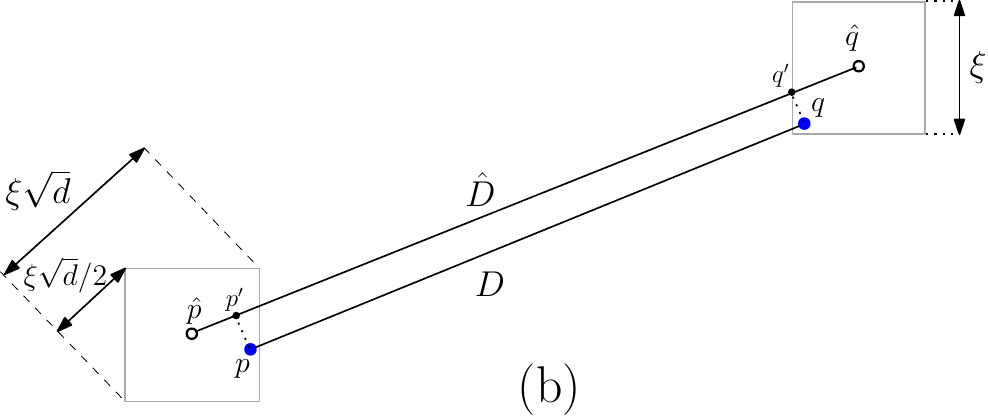}
       \caption{Two cases in proof of the Theorem 2. Two central-cell points $\hat{p}$ and $\hat{q}$ are used for computing the approximate diameter $\hat{D}$. Their corresponding true points are $p$ and $q$.}
       \label{fig:Lem}                   
    \end{center}
\end{figure}
\vspace{0.25cm}
For first case (Figure~\ref{fig:Lem} (a)), we can rotate line $\hat{p}\hat{q}$ such that two points $\hat{p}$ and $\hat{q}$ are projected on line $pq$. Let these two projected points be $\hat{p}'$ and $\hat{q}'$, and we set $d_{1}=||p-\hat{p}'||$ and $d_{2}=||q-\hat{q}'||$. We know that the side length of each cell in a grid which is used for point set $\hat{\mathcal{S}}$ is $\xi$. So, the hypercube (cell) diagonal is $\xi\sqrt{d}$. From Figure~\ref{fig:Lem} (a) it can be found that $d_{1}< \xi\sqrt{d} /2$ and $d_{2}< \xi\sqrt{d} /2$. Therefore, we have 
$$D=\hat{D}+d_{1}+d_{2},$$
$$D \leqslant \hat{D}+\xi\sqrt{d} /2+\xi\sqrt{d} /2,$$
$$D \leqslant \hat{D}+\xi\sqrt{d},$$
\begin{equation}
D-\xi\sqrt{d}\leqslant \hat{D}.
\end{equation}
Similarly, for the second case (Figure~\ref{fig:Lem} (b)), we can project two points $p$ and $q$ on line $\hat{p}\hat{q}$. Let these two projected points be $p'$ and $q'$. We know that $c_{1}=||\hat{p}-p'||<\xi\sqrt{d} /2$ and $c_{2}=||\hat{q}-q'||<\xi\sqrt{d} /2$. Therefore, we have
$$\hat{D}=D+c_{1}+c_{2},$$ 
$$\hat{D} \leqslant D+\xi\sqrt{d} /2+\xi\sqrt{d} /2,$$
\begin{equation}
\hat{D} \leqslant D+\xi\sqrt{d}.
\end{equation}
Then, from (5) and (6) we can result:
\begin{equation}
D-\xi\sqrt{d} \leqslant \hat{D}\leqslant  D+\xi\sqrt{d}.
\end{equation}
Since we know that $\xi=\varepsilon \ell /2\sqrt{d}$, we have:
\begin{equation}
D-\varepsilon \ell/2 \leqslant \hat{D}\leqslant D+ \varepsilon \ell/2.
\end{equation}
Now, we can simplify (8) as following:
\begin{equation}
D \leqslant \hat{D}+\varepsilon \ell/2 \leqslant D+\varepsilon \ell.
\end{equation}
We know that $\ell \leqslant D$. For this reason we can result:
\begin{equation}
D \leqslant \hat{D}+\varepsilon \ell/2 \leqslant (1+\varepsilon)D.
\end{equation} 
Finally, if we assume that $\tilde{D}=\hat{D}+\varepsilon \ell/2$, we have:
\begin{equation}
D \leqslant \tilde{D} \leqslant (1+\varepsilon)D.
\end{equation}
Therefore, the theorem is proven.  
\end{proof}
\subsection{The modified algorithm}
In this subsection, we present a modified version of our proposed algorithm by combining it with a recursive approach due to Chan~\cite{Cha02}. Hence, we first explain Chan's recursive approach and then use it in a phase of our proposed algorithm. As mentioned before, Agarwal et al.~\cite{AMS92} proposed a $(1 +\varepsilon)$-approximation algorithm for computing the diameter of a set of $n$ points in $\mathbb{R}^{d}$ with $O(n/\varepsilon^{(d-1)/2})$ running time by projecting on directions. In fact, they found a small set of directions which can approximate well all directions. This can be done by forming unit vectors which start from origin to grid-points of a uniform grid on a unit sphere~\cite{AMS92}, or to grid-points on the boundary of a box~\cite{Cha06}. These sets of directions have cardinality $O(1/\varepsilon^{(d-1)/2})$. The following observation explains how we can find these directions on the boundary of a box.
\newtheorem{Ob1} [enumi] {Observation}{\bfseries}{\itshape} 
\begin{Ob1}
(\cite{Cha06}) Consider a box $B$ which includes origin $o$ such that the boundary of this box ($\partial B$) be in distance at least 1 from origin. For a $\sqrt{\varepsilon/2}$-grid on $\partial B$ and for each vector $\vec{x}$, there is a grid point $\vec{a}$ on $\partial B$ such that the angle between two vectors $\vec{a}$ and $\vec{x}$ is at most  $arccos(1-\varepsilon/8)\leqslant \sqrt{\varepsilon}$. 
\end{Ob1}
\begin{proof} By scaling, we may assume that $x \in \partial B$. Since, in a $\sqrt{\varepsilon/2}$-grid the diagonal of each cell is $\sqrt{\varepsilon}$, so there is a grid point $a$ such that: $||a-x||\leqslant \sqrt{\varepsilon}/2$. Then, we have:
$$\qquad \quad 2a\cdot x\geqslant||a||^{2}+||x||^{2}-(\sqrt{\varepsilon})^{2}/4,$$
$$\qquad \geqslant 2||a||||x||-\varepsilon/4,$$
\begin{equation}
\qquad \quad \geqslant 2||a||||x||(1-\varepsilon/8).
\end{equation}
This results the observation, because: $cos \angle (a,x)=a\cdot x/ ||a|| ||x||$. 
\end{proof}

\begin{figure}
    \begin{center}
       \includegraphics [height=3.95 cm]{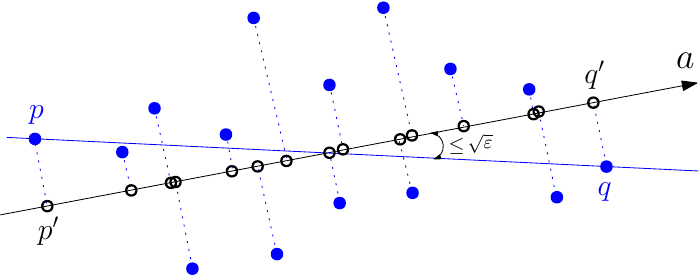}
       \caption{Projecting a point set on a direction $a$. True diametrical pair $(p,q)$ and its projected diametrical pair $(p',q')$.}
       \label{fig:Project}                    
    \end{center}
\end{figure}

This observation explains that grid-points on the boundary of a box ($\partial B$) form a set $V_{d}$ of $O(1/\varepsilon^{(d-1)/2})$ numbers of unit vectors in $\mathbb{R}^{d}$ such that for each $x \in \mathbb{R}^{d}$, there is a vector $a \in V_{d}$ from origin $o$ to a grid-point $a$ on $\partial B$, where the angle between two vectors $x$ and $a$ is at most $\sqrt{\varepsilon}$. On the other hand, according to observation 3, there is a vector $a \in V_{d}$ such that if $\alpha$ be the angle between two vectors $x$ and $a$, then $\alpha\leqslant arccos(1-\varepsilon/8)$, and so $cos \alpha \geqslant (1-\varepsilon/8)$. If $x'$ be the projection of the vector $x$ on the vector $a$, then:
$$||x||=\frac{||x'||}{cos \alpha}$$
$$ \qquad \qquad \quad \leqslant ||x'|| \frac{1}{(1-\frac{\varepsilon}{8})}$$
$$ \qquad  \qquad  \qquad \qquad \qquad \qquad \leqslant ||x'||(1+\frac{\varepsilon}{8}+\frac{\varepsilon^{2}}{8^{2}}+\frac{\varepsilon^{3}}{8^{3}}+\cdots)$$  
\begin{equation}
\qquad \qquad \quad \leqslant ||x'||(1+\varepsilon).
\end{equation}
So, we have:
\begin{equation}
||x'||\leqslant ||x|| \leqslant (1+\varepsilon)||x'||.
\end{equation}
This means that if pair $(p,q)$ be the diametrical pair of a point set, then there is a vector $a \in V_{d}$ such that the angle between two vectors $pq$ and $a$ is at most $\sqrt{\varepsilon}$. See Figure~\ref{fig:Project}. Then, pair $(p',q')$ which is the projection of pair $(p,q)$ on the vector $a$, is a $(1+\varepsilon)$-approximation of the true diametrical pair $(p,q)$, and we have: \begin{equation}
||p'-q'||\leqslant ||p-q|| \leqslant (1+\varepsilon) ||p'-q'||.
\end{equation}
\vspace {0.05 cm} 
\vspace {-0.05 cm}

In other words, we can project point set $\mathcal{S}$ on $O(1/\varepsilon^{(d-1)/2})$ 
directions for all $a \in V_{d}$, and compute a $(1+\varepsilon)$-approximate of the diameter by finding maximum diameter between all directions. We project $n$ points on
$|V_{d}|=O(1/\varepsilon^{(d-1)/2})$ directions. In addition, computing the extreme points on each direction $a \in V_{d}$ takes $O(n)$ time. Consequently, Agarwal et al.~\cite{AMS92} algorithm computes a $(1+\varepsilon)$-approximate of the diameter in $O(n/\varepsilon^{(d-1)/2})$ time. Chan~\cite{Cha02} proposes that if we reduce number of points from $n$ to $O(1/\varepsilon^{d-1})$ by rounding to a grid and then apply Agarwal et al.~\cite{AMS92} method on this rounded point set, we need $O((1/\varepsilon^{d-1})/\varepsilon^{(d-1)/2})=O(1/\varepsilon^{3(d-1)/2})$ time to compute the maximum diameter over all $O(1/\varepsilon^{(d-1)/2})$ directions. Taking into account $O(n)$ time for rounding to a grid, this new approach takes $O(n+1/\varepsilon^{3(d-1)/2})$ time for computing a $(1+\varepsilon)$-approximation for the diameter of a set of $n$ points. Moreover, Chan~\cite{Cha02} observed that the bottleneck of this approach is the large number of projection operations. Hence, he proposes that we can project points on a set of $O(1/\sqrt{\varepsilon})$ 2-dimensional unit vectors instead of $O(1/\varepsilon^{(d-1)/2})$ $d$-dimensional unit vectors to reduce the problem to $O(1/\sqrt{\varepsilon})$ numbers of $(d-1)$-dimensional subproblems which can be solved recursively. Then, according to relation (14), for a vector  $x \in \mathbb{R}^{2}$, there is a vector $a$ such that:
\begin{equation}
||x'||\leqslant ||x|| \leqslant (1+\varepsilon)||x'||, \hspace*{0.3cm} x\in \mathbb{R}^{2}. 
\end{equation}
Since $a$ is a unit vector ($||a||=1$), therefore, $||x'||=(a\cdot x)/||a||=a\cdot x$. Hence, we can rewrite the previous relation as following:
\begin{equation}
(a\cdot x)^{2}\leqslant ||x||^{2} \leqslant (1+\varepsilon)^{2}(a\cdot x)^{2}, \hspace*{0.1cm} \quad x\in \mathbb{R}^{2}, \quad a \in V_{2}, 
\end{equation}
or
\begin{equation}
(a_{1}x_{1}+a_{2}x_{2})^{2}\leqslant  (x_{1}^{2}+x_{2}^{2})\leqslant (1+\varepsilon)^{2}(a_{1}x_{1}+a_{2}x_{2})^{2}, \hspace*{0.3cm} a \in V_{2}. 
\end{equation}
when $x_{i}$ is the $i$th coordinate for a point $x \in \mathbb{R}^{d}$.
We can expand (18) to:
\begin{equation}
(a_{1}x_{1}+a_{2}x_{2})^{2}+\cdots+x_{d}^{2}\leqslant (x_{1}^{2}+x_{2}^{2}+\cdots+x_{d}^{2}) \leqslant (1+\varepsilon)^{2}((a_{1}x_{1}+a_{2}x_{2})^{2}+\cdots+x_{d}^{2}).
\end{equation}
Now, define the projection $\pi_{a}:\mathbb{R}^{d}\rightarrow \mathbb{R}^{d-1}: \pi_{a}(x)=(a_{1}x_{1}+a_{2}x_{2},x_{3},\cdots,x_{d}) \in \mathbb{R}^{d-1}$. Then, we can rewrite relation (19) for each vector $x\in \mathbb{R}^{d}$ as following:
\begin{equation}
||\pi_{a}(x)||^{2} \leqslant ||x||^{2}  \leqslant (1+\varepsilon)^{2}||\pi_{a}(x)||^{2}, \hspace*{0.2cm} a \in V_{2}. 
\end{equation}
So, since $||\pi_{a}(p-q)||=||\pi_{a}(p)||-||\pi_{a}(q)||$ we have for diametrical pair $(p,q)$:
\begin{equation}
||\pi_{a}(p-q)|| \leqslant ||p-q|| \leqslant (1+\varepsilon)||\pi_{a}(p-q)||, \hspace*{0.2cm} a \in V_{2}. 
\end{equation}
Therefore, for finding a $(1+O(\varepsilon))$-approximation for the diameter of point set $P\subseteq \mathbb{R}^{d}$, it is sufficient that we approximate recursively the diameter of a projected point set $\pi_{a}(P)\subset \mathbb{R}^{d-1}$ over each of the vectors $a \in V_{2}$. Then, the maximum diametrical pair computed in the recursive calls is a $(1+O(\varepsilon))$--approximation to the
diametrical pair. Now, let us reduce the number of points from $n$ to $O(1/\varepsilon^{d-1})$ by rounding to a grid. Let we denote the required time for computing the diameter of $m$ points in
$d$-dimensional space with $t_{d}(m)$, then for a rounded point set on a grid with $m=O(1/\varepsilon^{d-1})$ points, this approach breaks the
problem into $O(1/\sqrt{\varepsilon})$
subproblems in a $(d-1)$ dimension. Hence, we have a recurrence    
$t_{d}(m)=O(m+1/\sqrt{\varepsilon}t_{d-1}(O(1/\varepsilon^{d-1})))$. By assuming $E = 1/\varepsilon$, we
can rewrite the recurrence as:
\begin{equation}
t_{d}(m)=O(m+E^{\frac{1}{2}}t_{d-1}(O(E^{d-1}))).
\end{equation} 
This can be solved to: $t_{d}(m)=O(m+E^{d-\frac{1}{2}})$.
In this case, $m=O(1/\varepsilon^{d-1})$, so, this recursive takes $O(1/\varepsilon^{d-\frac{1}{2}})$ time. Taking into account $O(n)$ time, we spent for rounding to a grid at the first, Chan's recursive approach computes a $(1+O(\varepsilon))$-approximation for the diameter of a set of $n$ points in $O(n+1/\varepsilon^{d-\frac{1}{2}})$ time~\cite{Cha02}.
 
In the following, we use Chan's recursive approach in a phase of our proposed algorithm and present a modified version of it with running time $O(n+1/\varepsilon^{\frac{2d}{3}-\frac{1}{3}})$.
  
\smallskip‎
\hrule‎
\\
\textbf{ Algorithm 2‎:} APPROXIMATE DIAMETER 2 $ (\mathcal{S},\varepsilon)$ 
\hrule‎
\\ 
\footnotesize‎ 
\smallskip‎
\textbf{‎Input‎:} a set $\mathcal{S}$ of $n$ points in $\mathbb{R}^{d}$ and an error parameter $\varepsilon$.

\hspace*{-0.4cm}\textbf{‎Output‎:} approximate diameter $\tilde{D}$. 

\hspace{-0.4 cm} 1: \hspace{0.03cm} Compute the axis-parallel bounding box $B(\mathcal{S})$ for a point set $\mathcal{S}$.

\hspace{-0.4 cm} 2: \hspace{0.03cm} $\ell \leftarrow$ Find the length of the largest side in $B(\mathcal{S})$. 

\hspace{-0.4 cm} 3: \hspace{0.03cm} Set $\xi\leftarrow \varepsilon \ell/2\sqrt{d}$ and $\xi_{2}\leftarrow \varepsilon^{\frac{1}{3}} \ell/2\sqrt{d}$.

\hspace{-0.4 cm} 4: \hspace{0.03cm} $\hat{\mathcal{S}}\leftarrow$ Round each point of $\mathcal{S}$ to its central-cell point in a $\xi$-grid.

\hspace{-0.4 cm} 5: \hspace{0.03cm} $\hat{\mathcal{S}_{1}}\leftarrow$ Round each point of $\hat{\mathcal{S}}$ to its nearest grid-point in a $\xi_{2}$-grid.

\hspace{-0.4 cm} 6: \hspace{0.03cm} $\hat{D}_{1}\leftarrow$ Compute the diameter of the point set $\hat{\mathcal{S}_{1}}$ by brute-force, and simultaneously, \\ \hspace*{1.72cm} a list of the diametrical pair $(\hat{p}_{1},\hat{q}_{1})$, such that $\hat{D}_{1}=||\hat{p}_{1}-\hat{q}_{1}||$.

\hspace{-0.4 cm} 7: \hspace{0.03cm} Find points of $\hat{\mathcal{S}}$ which are in two hypercubes $\mathcal{B}_{1}=\mathcal{B}_{2\xi_{2}}(\hat{p}_{1})$ and $\mathcal{B}_{2}=\mathcal{B}_{2\xi_{2}}(\hat{q}_{1})$ \\ \hspace*{0.86cm} for each diametrical pair $(\hat{p}_{1},\hat{q}_{1})$. 

\hspace{-0.4 cm} 8: \hspace{0.04cm} $\tilde{D}\leftarrow$ Compute  $Diam(\mathcal{B}_{1},\mathcal{B}_{2})$, corresponding to each diametrical pair $(\hat{p}_{1},\hat{q}_{1})$ \\ \hspace*{1.6cm} by using Chan's~\cite{Cha02} recursive approach and return the maximum value \\ \hspace*{1.6cm} $||p'-q'||$ over all of them.

\hspace{-0.4 cm} 9: \hspace{0.01cm} Output $\tilde{D}$.
\smallskip‎
\hrule‎
\rmfamily‎ 
\normalsize‎

Now we will analyze the running time and approximation factor of the Algorithm 2. 
\newtheorem{Th3} [enumi] {Theorem}{\bfseries}{\itshape}
\begin{Th3}
A $(1+O(\varepsilon))$-approximation for the diameter of a set of n points in $d$-dimensional Euclidean space can be computed in $O(n+1/\varepsilon^{\frac{2d}{3}-\frac{1}{3}})$ time, where $0 < \varepsilon \leqslant 1$.
\end{Th3}
\begin{proof} As it can be seen, lines 1 to 5 of the Algorithm 2 are the same as the Algorithm 1. We do rounding to grids twice and reach to a point set $\hat{\mathcal{S}_{1}}$ in $O(n)$ time. In this case, the number of points in rounded points set $\hat{\mathcal{S}_{1}}$ is at most: 
\begin{equation}
\dfrac{(\ell+\epsilon_{2})^{d}}{(\epsilon_{2})^{d}}=\left(\dfrac{\ell}{\varepsilon^{\frac{1}{3}} \ell/2\sqrt{d}}+1\right)^{d}=\left(\dfrac{2\sqrt{d}} {\varepsilon^{\frac{1}{3}}}+1\right)^{d}=O\left(\dfrac{(2\sqrt{d})^{d}}{\varepsilon^{\frac{d}{3}}}\right) .
\end{equation}
This can be reduced to $O((2\sqrt{d})^{d}/\varepsilon^{\frac{d}{3}-\frac{1}{3}})$, by keeping only highest and lowest points which are the same in their $(d-1)$ coordinates. So, for finding all diametrical pairs of the point set $\hat{\mathcal{S}_{1}}$, we can use the quadratic brute-force algorithm with $O((2\sqrt{d})^{d}/\varepsilon^{\frac{d}{3}-\frac{1}{3}})^{2})=O((2\sqrt{d})^{2d}/\varepsilon^{\frac{2d}{3}-\frac{2}{3}})$ time. Then, for each diametrical pair $(\hat{p}_{1},\hat{q}_{1})\in\hat{\mathcal{S}_{1}}$, we compute two sets $\mathcal{B}_{1}$ and $\mathcal{B}_{2}$ which include points of set $\hat{\mathcal{S}}$ which are inside two hypercubes $\mathcal{B}_{2\xi_{2}}(\hat{p}_{1})$ and $\mathcal{B}_{2\xi_{2}}(\hat{q}_{1})$, respectively. Moreover, the number of points in two sets $\mathcal{B}_{1}$ or $\mathcal{B}_{2}$ is at most
\begin{equation}
\dfrac{Vol(\mathcal{B}_{2\xi_{2}})}{Vol(\mathcal{B}_{\xi})}=\dfrac{(2\varepsilon^{\frac{1}{3}} \ell/ 2\sqrt{d})^{d}}{(\varepsilon \ell/ 2\sqrt{d})^{d}}=\dfrac{(2\varepsilon^{\frac{1}{3}})^{d}} {\varepsilon^{d}}
=\dfrac{(2)^{d}}{\varepsilon^{\frac{2d}{3}}}.
\end{equation}
This can be reduced to $O((2)^{d}/\varepsilon^{\frac{2d}{3}-\frac{1}{3}})$, by keeping only highest and lowest points which are the same in their $(d-1)$ coordinates. Now, for computing $Diam(\mathcal{B}_{1},\mathcal{B}_{2})$, we use Chan's~\cite{Cha02} recursive approach instead of using the quadratic brute-force algorithm on the point set $\mathcal{B}_{1}\cup \mathcal{B}_{2}$. On the other hand, computing the diameter on a set of $O(1/\varepsilon^{\frac{2d}{3}-\frac{1}{3}})$ points using Chan's recursive approach takes the following recurrence based on relation (22): $t_{d}(m)=O(m+1/\sqrt{\varepsilon}t_{d-1}(O(1/\varepsilon^{\frac{2d}{3}-\frac{1}{3}})))$.
By assuming $E = 1/\varepsilon$, we
can rewrite the recurrence as:
\begin{equation}
t_{d}(m)=O(m+E^{\frac{1}{2}}t_{d-1}(O(E^{\frac{2d}{3}-\frac{1}{3}}))).
\end{equation} 
This can be solved to: $t_{d}(m)=O(m+E^{\frac{2d}{3}-\frac{1}{2}})$. In this case, $m=O(E^{\frac{2d}{3}-\frac{1}{3}})$, so, this recursive takes $O(E^{\frac{2d}{3}-\frac{1}{3}}+E^{\frac{2d}{3}-\frac{1}{2}})=O(1/\varepsilon^{\frac{2d}{3}-\frac{1}{3}})$ time. Also, if we have more than one diametrical pair $(\hat{p}_{1},\hat{q}_{1})$ in point set $\hat{\mathcal{S}_{1}}$, then this step takes at most $O( (2^{d})(2)^{d}/\varepsilon^{\frac{2d}{3}-\frac{1}{3}})=O( 2^{2d}/\varepsilon^{\frac{2d}{3}-\frac{1}{3}})$ time. Therefore, we can write the complexity time of the algorithm as following: 
$$T_{d}(n)=O(dn)+O(dn)+O\left(\dfrac{ (2\sqrt{d})^{2d}}{\varepsilon^{\frac{2d}{3}-\frac{2}{3}}}\right)+O(2^{d}dn)+O\left(\dfrac{2^{2d}}{\varepsilon^{\frac{2d}{3}-\frac{1}{3}}}\right),$$
$$\quad \hspace{0.2cm} \leqslant O(2^{d}dn)+O\left( \dfrac{(2\sqrt{d})^{2d}}{\varepsilon^{\frac{2d}{3}-\frac{1}{3}}}\right),$$
\begin{equation}
= O\left(2^{d}dn+ \dfrac{(2\sqrt{d})^{2d}}{\varepsilon^{\frac{2d}{3}-\frac{1}{3}}}\right).
\end{equation}
Since $d$ is fixed, we have:
\begin{equation}
T_{d}(n)=O\left(n+\dfrac{1}{\varepsilon^{\frac{2d}{3}-\frac{1}{3}}}\right).
\end{equation}
In addition, Chan's recursive approach in line 8 of the Algorithm 2 returns a diametrical pair $(p',q')$ which is a $(1+O(\varepsilon))$-approximation for the diametrical pair $(\hat{p},\hat{q}) \in \hat{\mathcal{S}}$. This means that:
\begin{equation}
||p'-q'||\leqslant||\hat{p}-\hat{q}||\leqslant (1+O(\varepsilon)) ||p'-q'||.
\end{equation}
Moreover, the diametrical pair $(\hat{p},\hat{q})$ is an approximation of the true diametrical pair $(p,q)\in \mathcal{S}$, and according to relation (10), we have:
\begin{equation}
||p-q||\leqslant||\hat{p}-\hat{q}||+\varepsilon \ell/2\leqslant (1+\varepsilon) ||p-q||.
\end{equation}
Hence, from (28) and (29) we can result:
$$||p-q||\leqslant||\hat{p}-\hat{q}||+
\varepsilon \ell/2,$$
$$\hspace*{0.2cm}\qquad \qquad \quad   \leqslant||\hat{p}-\hat{q}||+
\varepsilon ||\hat{p}-\hat{q}||,$$
$$\hspace*{0.2cm}\qquad \quad  \leqslant (1+\varepsilon) ||\hat{p}-\hat{q}||,$$
$$\hspace*{0.2cm}\qquad \qquad  \qquad \qquad \quad \leqslant (1+\varepsilon)((1+O(\varepsilon)) ||p'-q'||),$$
\begin{equation}
\qquad \qquad \qquad \leqslant (1+O(\varepsilon)) ||p'-q'||.
\end{equation}
So, Algorithm 2 finds a $(1+O(\varepsilon))$-approximation of the diameter of a point set $\mathcal{S}$ of $n$ points in $O(n+1/\varepsilon^{\frac{2d}{3}-\frac{1}{3}})$ time. Therefore, this completes the proof.
\end{proof}

\section{Conclusion}
We have presented a new $(1+\varepsilon)$-approximation algorithm to compute the diameter of a point set $\mathcal{S}$ of $n$ points in $\mathbb{R}^{d}$ for a fixed dimension $d$ in $O(n+ 1/\varepsilon^{d-1})$ time, where $0 <\varepsilon\leqslant 1$. Moreover, we show that the proposed algorithm can be modified to a $(1+O(\varepsilon))$-approximation algorithm with $O(n+ 1/\varepsilon^{\frac{2d}{3}-\frac{1}{3}})$ time. Our proposed algorithms provide some improvements in comparison with existing algorithms in terms of simplicity, understanding and data structure.

%\bibliographystyle{unsrt}
%\bibliography{paper}
\end{document}